\documentclass[twocolumn,PRB,showpacs,preprintnumbers,amsmath,amssymb]{revtex4}

\usepackage{graphicx}
\usepackage{dcolumn}
\usepackage{bm}

\begin{document}

\title{Magnetic ordering and dense Kondo behavior in
EuFe$_{2}$P$_{2}$}

\author{Chunmu Feng}
\affiliation{Department of Physics, Zhejiang University, Hangzhou
310027, China} \affiliation{Test and Analysis Center, Zhejiang
University, Hangzhou 310027, China}

\author{Zhi Ren, Shenggao Xu, Shuai Jiang, Zhu'an Xu}
\affiliation{Department of Physics, Zhejiang University, Hangzhou
310027, China} \affiliation{State Key Lab of Silicon Materials,
Zhejiang University, Hangzhou 310027, China}

\author{Guanghan Cao}
\email[corresponding author: ]{ghcao@zju.edu.cn}
\affiliation{Department of Physics, Zhejiang University, Hangzhou
310027, China} \affiliation{State Key Lab of Silicon Materials,
Zhejiang University, Hangzhou 310027, China}

\author{I. Nowik, I. Felner}
\affiliation{Racah Institute of Physics, The Hebrew University,
Jerusalem 91904, Israel}

\author{Kazuyuki Matsubayashi, Yoshiya Uwatoko}
\affiliation{Institute for Solid State Physics, The University of
Tokyo, Kashiwanoha, Kashiwa, Chiba 277-8581, Japan}

\begin{abstract}
Ternary iron phosphide EuFe$_2$P$_2$ with ThCr$_2$Si$_2$-type
structure has been systematically studied by the measurements of
crystal structure, magnetization, M\"{o}ssbauer effect, transport
properties and specific heat. The structural refinement result
confirms no direct P-P covalent bonding. The M\"{o}ssbauer spectra
indicate no magnetic moment for the Fe atoms and, that the Eu ions
are divalent in the whole temperatures. The Eu$^{2+}$ spins order
ferromagnetically at $T_C$=29 K, followed by a possible helimagnetic
ordering below $T_{HM}$=26 K, where the Eu$^{2+}$ moments tilt a
little from the $c$-axis. External magnetic field increases the
$T_C$ gradually, but suppresses the $T_{HM}$ rapidly.
(Magneto)resistivity data indicate characteristic dense Kondo
behavior above the Curie temperature. The result is discussed in
terms of the interplay between intersite RKKY and intrasite Kondo
interactions.
\end{abstract}

\pacs{75.50.Cc, 75.30.Cr, 75.30.Mb, 76.80.+y}


\maketitle

\section{\label{sec:level1}Introduction}

The interplay between 4$f$ and conduction electrons in intermetallic
compounds has led to a wide variety of novel ground
states,\cite{Radousky,Stewart} attracting sustained interest in
condensed matter physics community. In the pnictide family,
Eu$T$$_{2}$$Pn$$_{2}$ ($T$=transition metals; $Pn$=As or P) offers
us a rare opportunity to access such an interplay. The ternary
compound crystallizes in ThCr$_2$Si$_2$-type structure, consisting
of Eu-sublattice with 4$f$ electrons and $T$-sublattice with 3$d$
electrons. Europium is known as a special rare-earth element due to
the two stable valence configurations: Eu$^{2+}$ and Eu$^{3+}$,
showing a large moment ($J$=$S$=7/2) and zero moment ($J$=0),
respectively. In most cases, europium shows the lower valence with
high magnetic moment, which renders magnetically ordered ground
states. However, mixed valence state for Eu was evidenced by
M\"{o}ssbauer investigations in a "collapsed" phase
EuNi$_{2}$P$_{2}$.\cite{JPCS1988} By applying pressures, a
structural transition toward the collapsed phase was observed in
EuCo$_{2}$P$_{2}$ and EuFe$_{2}$P$_{2}$, accompanying with a partial
valence transition.\cite{PB1998, PRB2001} In earlier studies,
valence fluctuations of Eu were also demonstrated in
EuCu$_{2}$Si$_{2}$ system.\cite{PRL1973,PRL1982}

The europium iron pnictide EuFe$_{2}$$Pn$$_{2}$, first synthesized
more than 30 years ago,\cite{JSSC1978} exhibits totally different
physical properties for $Pn$=As and P. EuFe$_{2}$As$_{2}$ undergoes
a spin-density-wave (SDW) transition in the Fe sublattice at 200 K,
followed by an AFM ordering of Eu$^{2+}$ moments at 20
K.\cite{Raffius,Ren2008,JOP2008} By contrast, as reported in Ref.
~\cite{JPCS1988}, the Fe atoms do not carry local moments while the
Eu$^{2+}$ spins order ferromagnetically at 27 K in
EuFe$_{2}$P$_{2}$. Surprisingly, by doping P into
EuFe$_{2}$As$_{2}$, both superconductivity coming from Fe 3$d$
electrons and ferromagnetism due to Eu 4$f$ moments were observed in
EuFe$_2$(As$_{0.7}$P$_{0.3}$)$_2$.\cite{Ren2009}

While EuFe$_{2}$As$_{2}$ has been extensively studied
recently,\cite{Ren2008,JOP2008,Jeevan2008,Jiang2009,Wu2009,hp1,hp2,RXS,neutron}
few works\cite{JPCS1988,PB1998,PRB2001} have been devoted to
EuFe$_{2}$P$_{2}$. To the best of our knowledge, the transport and
thermodynamic properties of EuFe$_{2}$P$_{2}$ have not been reported
so far. Moreover, the contrasting behaviors between an iron arsenide
and its sister phosphide are explicitly demonstrated in CeFe$Pn$O
system: CeFeAsO serves as a parent compound for high temperature
superconductors,\cite{ChenGF} but CeFePO has been recognized as a
heavy Fermion metal with ferromagnetic correlation.\cite{CeFePO}
Therefore, what EuFe$_{2}$P$_{2}$ behaves is an important issue to
be investigated. In this paper, we performed a systematic study on
EuFe$_{2}$P$_{2}$ by the measurements of crystal structure,
transport properties, specific heat, as well as magnetic properties
and M\"{o}ssbauer spectra. The Eu valence state is confirmed to be
2+ in the whole temperatures, and the Eu$^{2+}$ moments order in a
complex manner rather than the simple reported ferromagnetism at low
temperatures. Strikingly, EuFe$_{2}$P$_{2}$ shows a dense Kondo
behavior. Our result demonstrates that, as an Eu-containing
compound, EuFe$_{2}$P$_{2}$ sets a rare example displaying the
interplay between Kondo and Ruderman-Kittel-Kasuya-Yosida (RKKY)
interactions.

\section{\label{sec:level1}Experimental details}

Polycrystalline samples of EuFe$_2$P$_2$ were synthesized by solid
state reaction between EuP and Fe$_{2}$P, as reported
previously.\cite{Ren2009} EuP was presynthesized by heating europium
grains and phosphorus powders very slowly to 1173 K, then holding
for 36 h. Fe$_{2}$P was presynthesized by reacting iron and
phosphorus powders at 973 K for 24 h from stoichiometric amounts of
the elements. All the starting materials have the purity better than
99.9 \%. Powders of EuP and Fe$_{2}$P were weighed according to the
stoichiometric ratio, thoroughly ground and pressed into pellets in
an argon-filled glove-box. The pellets were then sealed in an
evacuated quartz tube and sintered at 1273 K for 36 h then cooled
slowly to room temperature.

Powder X-ray diffraction (XRD) was carried out using a D/Max-rA
diffractometer with Cu K$_{\alpha}$ radiation and a graphite
monochromator. The structural refinement was performed using the
Programme RIETAN 2000.\cite{Rietan} The electrical resistivity was
measured using a standard four-probe method. Thermoelectric power
measurements were carried out by a steady-state technique with a
temperature gradient $\sim$ 1 K/cm. Magnetoresistance (MR) and
specific heat measurements were performed on a Quantum Design
Physical Property Measurement System (PPMS-9). The dc magnetization
was measured on a Quantum Design Magnetic Property Measurement
System (MPMS-5).

M\"{o}ssbauer studies were performed using a conventional constant
acceleration drive. The sources were 50 mCi $^{57}$Co:Rh for
$^{57}$Fe spectra and 200 mCi $^{151}$Sm$_2$O$_3$ source for the
$^{151}$Eu spectra. The absorbers were measured in a Janis model
SHI-850-5 closed cycle refrigerator. The spectra were analyzed in
terms of least square fit procedures to theoretical expected
spectra, including full diagonalization of the hyperfine interaction
spin Hamiltonian. The analysis of the $^{151}$Eu spectra considered
also the exact shape of the source emission line, as shown in
Ref.~\cite{ME-1}. The velocity calibration was performed with an
$\alpha$-iron foil at room temperature. The reported isomer shift
(I.S.) values for iron are relative to the Fe foil, for europium
relative to the oxide source at room temperature.

\section{\label{sec:level1}Results and Discussion}

\subsection{\label{sec:level2}Crystal Structure}

Figure 1 shows the powder XRD pattern for the as-prepared
EuFe$_2$P$_2$ sample. No obvious secondary phase can be detected. By
employing ThCr$_2$Si$_2$-type structure with the space group of
I$4/mmm$, the crystal structure were successfully refined (the
reliable factor $R_{wp}$=0.098; the goodness of fit is 1.61). The
fitted lattice parameters are $a$=3.8178(1) {\AA} and $c$=11.2372(3)
{\AA}, in agreement with the literature values [$a$=3.818(1) {\AA}
and $c$=11.224(4) {\AA}].\cite{JSSC1978} Compared with the
counterpart of arsenide EuFe$_{2}$As$_{2}$,\cite{JOP2008} the $a$
and $c$ axes decrease by 2.3 \% and 7.3 \%, respectively. The larger
decrease in $c$ axis suggests stronger interlayer coupling.

\begin{figure}
\includegraphics[width=7.5cm]{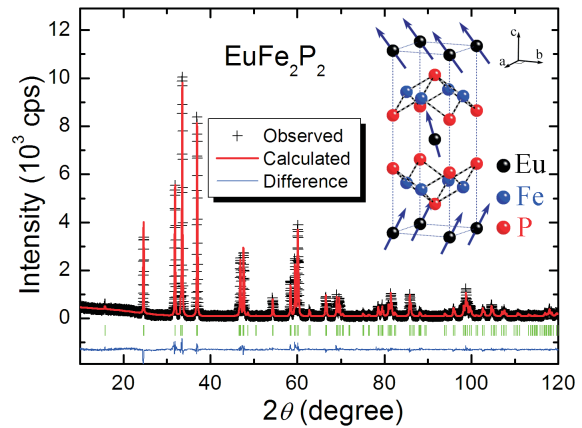}
\caption{(Color online) Rietveld refinement profile of powder X-ray
diffraction at room temperature for EuFe$_2$P$_2$. The inset shows
the crystal structure. A possible magnetic structure is also
illustrated based on the magnetization and M\"{o}ssbauer results.}
\end{figure}

Detailed structural comparison between EuFe$_2$P$_2$ and
EuFe$_2$As$_2$ can be seen in Table 1. The position of phosphorus is
closer to the iron planes, which leads to 13.7 \% decrease in the
thickness of Fe$_{2}$$Pn$$_{2}$ layers. On the other hand, the
spacing of Fe$_{2}$$Pn$$_{2}$ layers, namely the $Pn-Pn$ distance,
only decreases by 2.1 \%. One note that the P-P distance (3.263
{\AA}) in EuFe$_2$P$_2$ is much larger than the threshold value of
$\sim$2.3 {\AA}\cite{JPC1985} for the P-P bonding along the $c$
axis. Therefore, unlike the collapsed phase
EuNi$_2$P$_2$,\cite{JPCS1988} there is no covalent P-P bonding in
EuFe$_2$P$_2$ at ambient condition. This result is consistent with
the previous report.\cite{PB1998} In the concept of bond valence
sum,\cite{bvs1} the formal valence of Eu is calculated to be 1.89
using the Eu$-$P bondlength data and the related
parameters.\cite{bvs2}

\begin{table}
\caption{\label{tab:table1}Comparison of room-temperature crystal
structures for EuFe$_{2}$As$_{2}$ (Ref. ~\cite{JOP2008}) and
EuFe$_{2}$P$_{2}$ (this work). The atomic coordinates of Eu, Fe and
$Pn$ are (0, 0, 0), (0.5, 0, 0.25) and (0, 0, $z$), respectively.}

\begin{tabular}{lcr}
\hline
Compounds&EuFe$_{2}$As$_{2}$&EuFe$_{2}$P$_{2}$\\
\hline space group &I$4/mmm$ &I$4/mmm$ \\
$a$ ({\AA}) & 3.9062(3) &3.8178(1)\\
$c$ ({\AA}) & 12.1247(2) &11.2372(3)\\
$V$ ({\AA}$^3$) & 185.01(1) & 163.79(1)\\
$z$ of $Pn$ & 0.3625(1) & 0.3548(2)\\
Fe$_2$$Pn$$_2$-layer thickness ({\AA}) & 2.728(2) & 2.355(2)\\
$Pn-Pn$ distance ({\AA}) & 3.333(2) & 3.263(2)\\
Eu-$Pn$ bondlength ({\AA}) & 3.226(2) & 3.154(2)\\
$Pn$-Fe-$Pn$ angle ($^{\circ}$) & 110.1(1) & 116.7(1)\\
Bond valence sum for Eu & 1.93 & 1.89\\
\hline
\end{tabular}
\end{table}

\subsection{\label{sec:level2}Magnetization}

Although the XRD experiment shows no obvious secondary phase, the
magnetic measurement indicates a step-like decrease of
susceptibility from 200 to 300 K (not shown here). Similar phenomena
was observed previously in the EuFe$_2$(As$_{0.7}$P$_{0.3}$)$_2$
sample,\cite{Ren2009} which is due to the presence of trace amount
of ferromagnetic impurity Fe$_2$P with a Curie point at 306
K.\cite{JPSJ1960} The molar fraction of Fe$_2$P was estimated to be
below 1\% from the $M(H)$ curves at 100 K. Figure 2(a) shows the
temperature dependence of magnetic susceptibility ($\chi$) below 120
K for the EuFe$_2$P$_2$ sample. The data of 35 K $\leq$$T$$\leq$ 120
K follows the modified Curie-Weiss law,
\begin{equation}
\chi=\chi_{0}+\frac{C}{T-\theta},
\end{equation}
where $\chi_{0}$ denotes the temperature-independent term, $C$ the
Curie-Weiss constant and $\theta$ the paramagnetic Curie
temperature. The fitted value of $\chi_{0}$ is as high as 0.22
emu/mol, which is mainly ascribed to the ferromagnetic Fe$_2$P
impurity. The fitting also yields the effective magnetic moments
$P_{eff}$=8.3 $\mu_{B}$ per formula unit and $\theta$=29 K. The
$P_{eff}$ value is consistent with the theoretical value of 7.94
$\mu_{B}$ for a free Eu$^{2+}$ ion (The slightly larger value is
also due to the influence of the tiny Fe$_2$P impurity). A
ferromagnetic transition is manifested by the rapid increase in
$\chi$ below 30 K, as well as the divergence of ZFC and FC data.
This result is basically consistent with the previous report
claiming ferromagnetic (FM) transition at 27 K by M\"{o}ssbauer and
magnetic susceptibility investigations.\cite{JPCS1988} However, the
Curie point ($T_C$) has 2-K difference. In fact, precise
determination of the $T_C$ by a single $M(T)$ curve is difficult
because of the large moments of Eu$^{2+}$. We thus measured series
of $M(H)$ curves nearby $T_C$. The data are shown in the plot of
$M^{2}$ vs. $H/M$ (so-called Arrot plot), which clearly indicates
that the Curie temperature is 29 K.

\begin{figure}
\includegraphics[width=7cm]{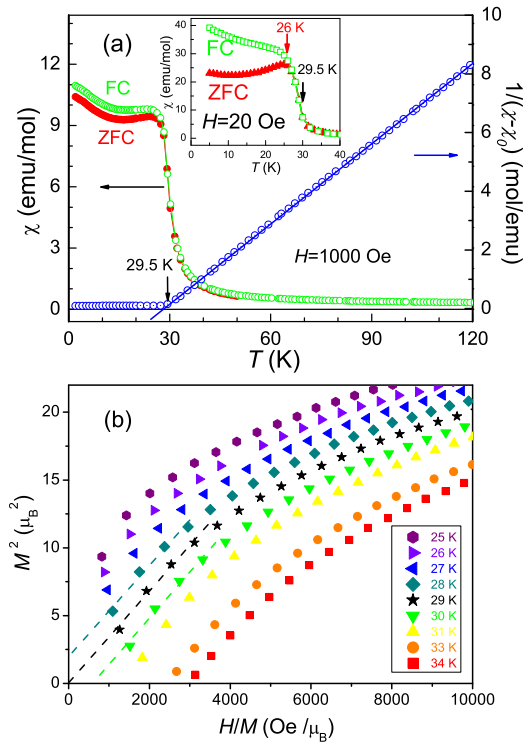}
\caption{(Color online) (a) Temperature dependence of magnetic
susceptibility measured under $H$=1 kOe for EuFe$_2$P$_2$. The
low-field susceptibility data ($H$=20 Oe) is shown in the inset. (b)
An Arrot plot for EuFe$_2$P$_2$.}
\end{figure}

Below the $T_C$, we note a kink at 26 K in the $\chi(T)$ data [shown
in the inset of figure 2(a)], which is quite different from those of
the conventional ferromagnet. The temperature dependence of
magnetization under various fields is displayed in figure 3. For
high magnetic fields, say $\mu_{0}H$=1 T, the magnetization
approximately saturates to the theoretical value of $gS$=7.0
$\mu_{B}$/f.u. at 2 K. In the case of low fields, however, there is
another magnetic transition below $T_C$, characterized by the
temperature-independent magnetization. This phenomenon is very much
similar to that in
Eu(Fe$_{0.89}$Co$_{0.11}$)$_2$As$_2$,\cite{Jiang-PRB2009} where a
helimagnetism was proposed. In the helimagnetic (HM) state, as
illustrated in the inset of figure 1, the Eu$^{2+}$ spins align
ferromagnetically within the Eu atomic planes, but the interlayer
spin directions differ in a fixed angle. Note that the $^{151}$Eu
M\"{o}ssbauer study below indicates that the spin direction tilts
about 20$^{\circ}$ from the $c$-axis, which gives rise to the
observed macroscopic ferromagnetism. External field suppresses the
$T_{HM}$ rapidly, but increases the $T_C$ gradually. The
stabilization of FM state by the external field was explained in our
previous paper.\cite{Jiang-PRB2009}

\begin{figure}
\includegraphics[width=7cm]{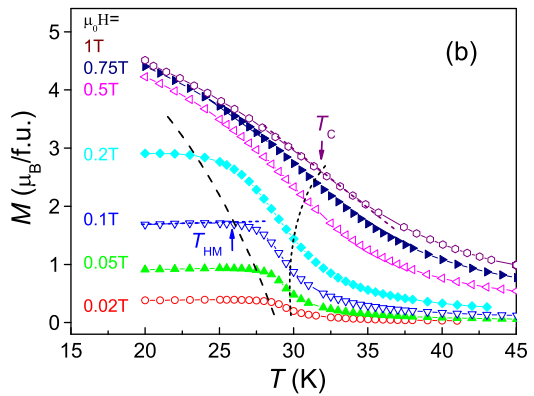}
\caption{(Color online) Temperature dependence of magnetization
under various magnetic fields for EuFe$_2$P$_2$. $T_C$ and $T_{HM}$
denote the ferromagnetic and helimagnetic transition temperatures,
respectively.}
\end{figure}

The dominant ferromagnetism in EuFe$_2$P$_2$ is further demonstrated
by the field-dependent magnetization, shown in figure 4. The
magnetization increases steeply with initially increasing $H$ and
tends to saturate for $H$$\geq$10$^{4}$ Oe. The saturated magnetic
moment is $\sim$6.7 $\mu_{B}$/f.u., close to the expected value of
7.0 $\mu_{B}$/f.u. In addition, a small hysteresis loop is presented
on closer examination. All these features are consistent with
basically ferromagnetic alignment of Eu$^{2+}$ moments.

\begin{figure}
\includegraphics[width=7cm]{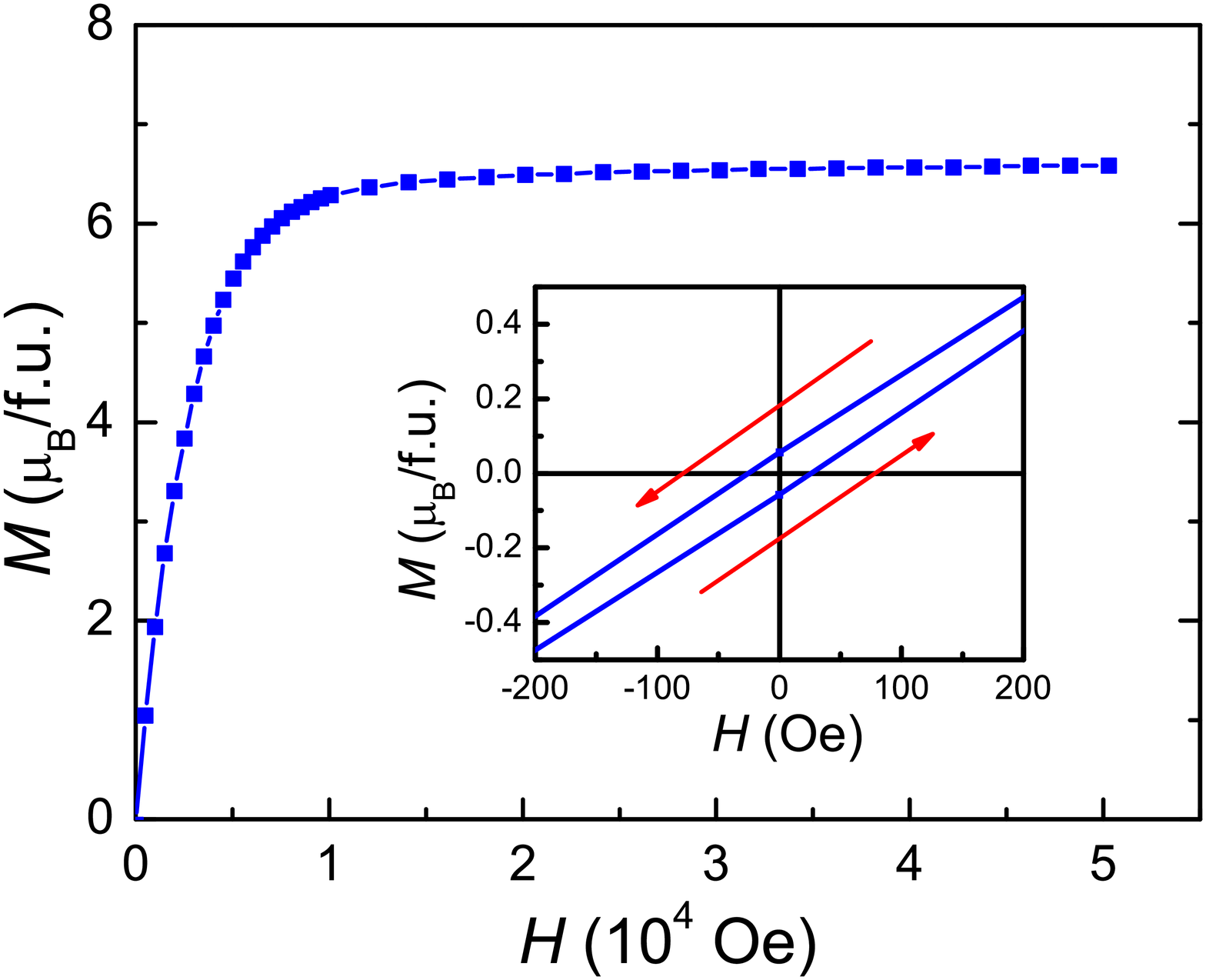}
\caption{(Color online) Field dependence of magnetization at 2 K for
EuFe$_2$P$_2$. The inset shows tiny magnetic hysteresis with the
coercive field of only 30 Oe.}
\end{figure}

\subsection{\label{sec:level2}M\"{o}ssbauer Spectra}

M\"{o}ssbauer Spectroscopy (MS) studies of $^{151}$Eu and $^{57}$Fe
in the system EuFe$_2$P$_2$ at temperatures 5 K to 297 K have been
performed. A previous study of this compound has been reported in
Ref. ~\cite{JPCS1988}.

$^{151}$Eu spectra [figure 5(a)] display pure quadrupole
interactions down to the magnetic ordering temperature of the Eu
sublattice. The values of the measured I.S. are $-$11.0(1) mm/s at
297 K, $-$11.4(1) mm/s at 40 K and $-$11.3(1) mm/s at 5 K, proving
that the Eu ions are divalent at all temperatures. The quadrupole
interaction values ($\frac{1}{4}e^{2}qQ_0$) are: $-$2.32(2) mm/s at
297 K, $-$2.55(2) mm/s at 40 K. At 5 K, the quadrupole shift is
$-$2.95(2) mm/s and the magnetic hyperfine field ($H_{eff}$) is
30.1(1) T. The quadrupole shift value at 5 K, when analyzed in the
approximation that the magnetic interactions are much larger than
the quadrupole interaction, indicates that $H_{eff}$ points along
the crystalline $c$-axis, the major axis of the axial electric field
gradient producing the quadrupole interaction. However, analyzing
the spectrum with a full diagonalization of the hyperfine Spin
Hamiltonian gives a better fit when the hyperfine field tilts away
from the $c$-axis by 20(5)$^{\circ}$. This is in contrast to
EuFe$_2$As$_2$ [figure 5(b)], where the Eu moments order
antiferromagnetically (of $A$-type) and the hyperfine field is
perpendicular to the $c$-axis. The same phenomenon is also observed
in Eu(Fe$_{0.9}$Ni$_{0.1}$)$_2$As$_2$,\cite{Felner2009} where the Eu
moment is ferromagnetically ordered.\cite{Ren-PRB2009} The observed
$H_{eff}$ in EuFe$_2$P$_2$ is higher than that ($H_{eff}$=26.2 T) in
EuFe$_2$As$_2$.

\begin{figure}
\includegraphics[width=7cm]{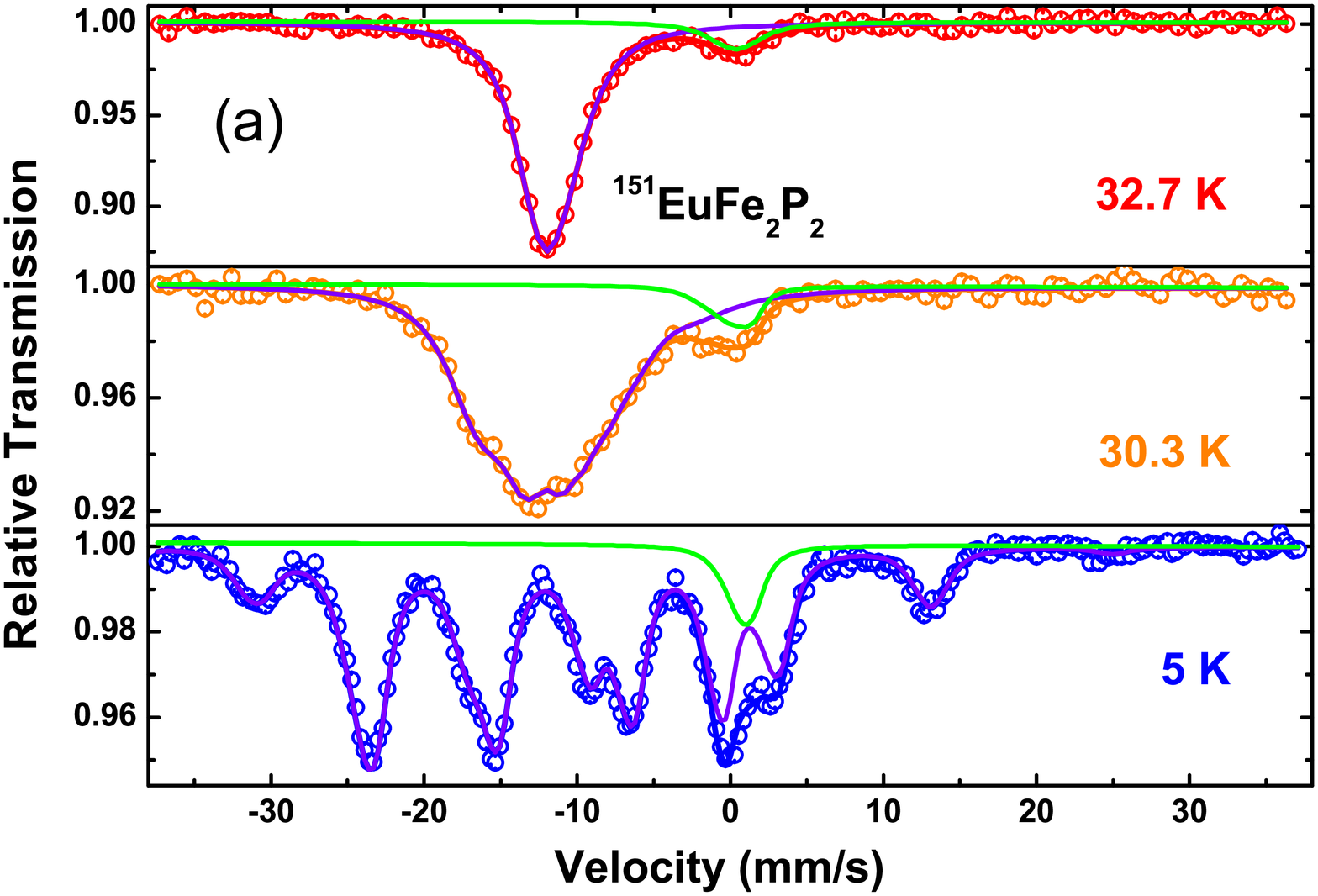}
\includegraphics[width=7cm]{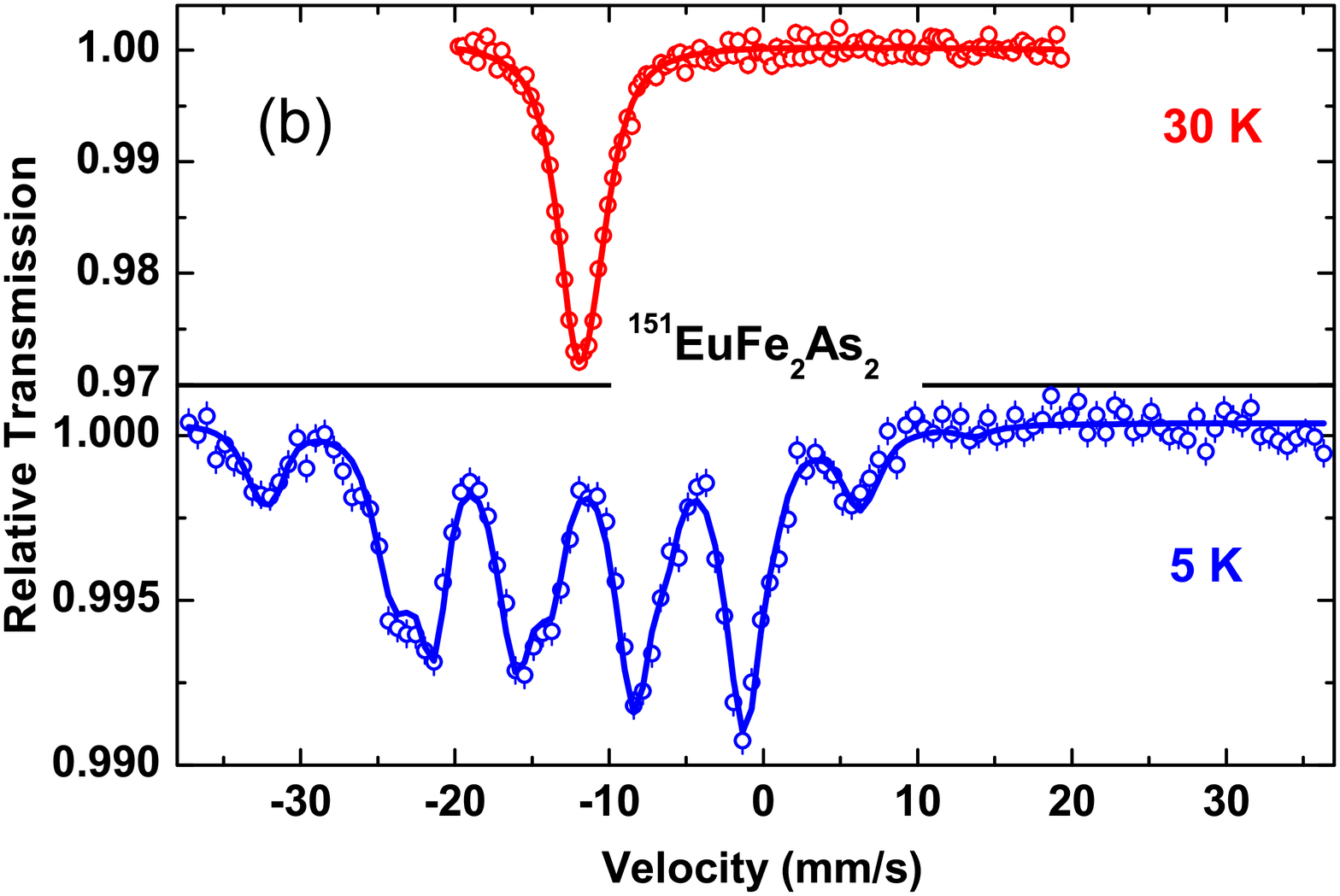}
\caption{(Color online) (a) $^{151}$Eu M\"{o}ssbauer spectra of
EuFe$_2$P$_2$, below (5 K), around (30.3 K) and above (32.7 K) the
ferromagnetic ordering temperature of the Eu sublattice. The
observed impurity site is probably due to minor Eu$_2$O$_3$ on the
surface of the grains. The hyperfine field is basically along the
$c$-axis. (b) $^{151}$Eu M\"{o}ssbauer spectra of EuFe$_2$As$_2$,
below (5 K) and above (30.0 K) the antiferromagnetic ordering
temperature of the Eu sublattice. The hyperfine field (26.2 T) is
perpendicular to the $c$-axic.}
\end{figure}

The $^{57}$Fe M\"{o}ssbauer spectra display a pure quadrupole
splitting down to $T_C$. The measured I.S. are: 0.28 mm/s at room
temperature, 0.38 mm/s at 34 K and 0.39 mm/s at 5 K. The quadrupole
interaction ($\frac{1}{4}e^{2}qQ_0$) values are: 0.16(1) mm/s at 297
K, 0.15(1) mm/s at 30 K and below $T_c$, at 5 K, 0.17(1) mm/s. A
small foreign phase (probably Fe$_2$P) of less than 5\% is also
present. In figure 6 one can observe the change in the spectrum
between 34 K to 5 K. No Fe magnetic moment is evidenced. At 5 K the
spectrum displays a small magnetic hyperfine field [$H_{eff}$
=0.97(2) T], acting on the iron nucleus. This small field is a
transferred field from the ferromagnetically ordered Eu sublattice,
as previously observed.\cite{JPCS1988} It was also observed in
ferromagnetic Eu(Fe$_{0.9}$Ni$_{0.1}$)$_2$As$_2$.\cite{Felner2009}
This transferred field seems to point along the $c$-axis, however
analyzing the spectrum with a full diagonalization of the hyperfine
Spin Hamiltonian produces a better fit when the $H_{eff}$ tilts from
the $c$-axis by 15(5)$^{\circ}$. Thus our measurements show, that
the transferred field on the iron site, points in the same direction
as that of the Eu magnetic moment.

\begin{figure}
\includegraphics[width=7cm]{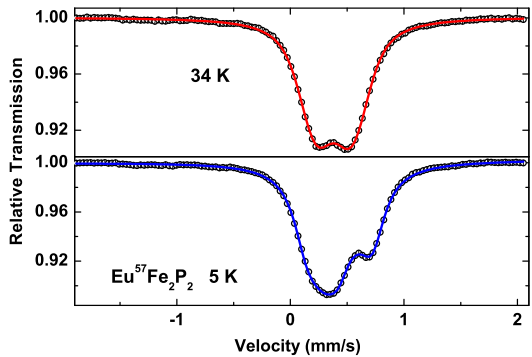}
\caption{(Color online) $^{57}$Fe M\"{o}ssbauer spectra of
EuFe$_2$P$_2$, below (5 K) and above (34 K) the ferromagnetic
ordering temperature of the Eu sublattice. The transferred magnetic
hyperfine field at 5 K, is along the $c$-axis.}
\end{figure}

\subsection{\label{sec:level2}Transport Properties}

\subsubsection{\label{sec:level3}Resistivity and Magnetoresistivity}

For intermetallic compounds, dense polycrystalline samples largely
exhibit intrinsic transport properties as single crystals do because
of well electrical contact between crystalline grains. Thus our
sample, which had metal luster by polishing, may reflect the
intrinsic transport properties. Figure 7(a) displays the temperature
dependence of resistivity ($\rho$) for EuFe$_2$P$_2$. The room
temperature resistivity is $\sim$0.15 m$\Omega$$\cdot$cm, which is
nearly the same to that of polycrystalline sample of
BaFe$_2$P$_2$,\cite{Jiang2009} but about one order of magnitude
smaller than the value of EuFe$_{2}$As$_{2}$.\cite{Ren2008} Unlike
EuFe$_{2}$As$_{2}$ which shows two anomalies in $\rho$ at 20 and 200
K, there is only one resistivity anomaly in EuFe$_2$P$_2$, i.e., a
kink at 29.2 K, corresponding to the aforementioned ferromagnetic
transition in Eu sublattice. The residual resistivity ratio (RRR),
defined as $\rho_{300K}$/$\rho_{2K}$, is $\sim$40, much reduced in
comparison with that of BaFe$_2$P$_2$($\sim$70).\cite{Jiang2009}
This result suggests additional magnetic scattering due to the
Eu$^{2+}$ moments in EuFe$_2$P$_2$. For \emph{simplicity}, we
roughly assume that the resistivity contribution from
electron-phonon scattering, denoted by $\rho_{e-ph}(T)$, is the same
for both materials. Then, the resistivity contribution from magnetic
scattering [$\rho_{mag}(T)$] in EuFe$_2$P$_2$ can be obtained simply
by a subtraction. As can be seen in figure 7(a), the $\rho_{mag}(T)$
data show a maximum at $\sim$55 K, which is reminiscence of dense
Kondo behavior in other systems such as Ce$T$Sb$_2$\cite{112-Kondo}
and CeNiGe$_3$ material.\cite{PRB2003-Kondo} It is noted that upon
applying a pressure up to 4 GPa, a broad resistivity peak appears
around 100 K without the subtraction of
$\rho_{e-ph}(T)$.\cite{Uwatoko} This pressure-enhanced Kondo effect
is very common,\cite{Thompson1985} because the 4$f$ level tends to
approach Fermi energy with increasing pressure.

\begin{figure}
\includegraphics[width=7cm]{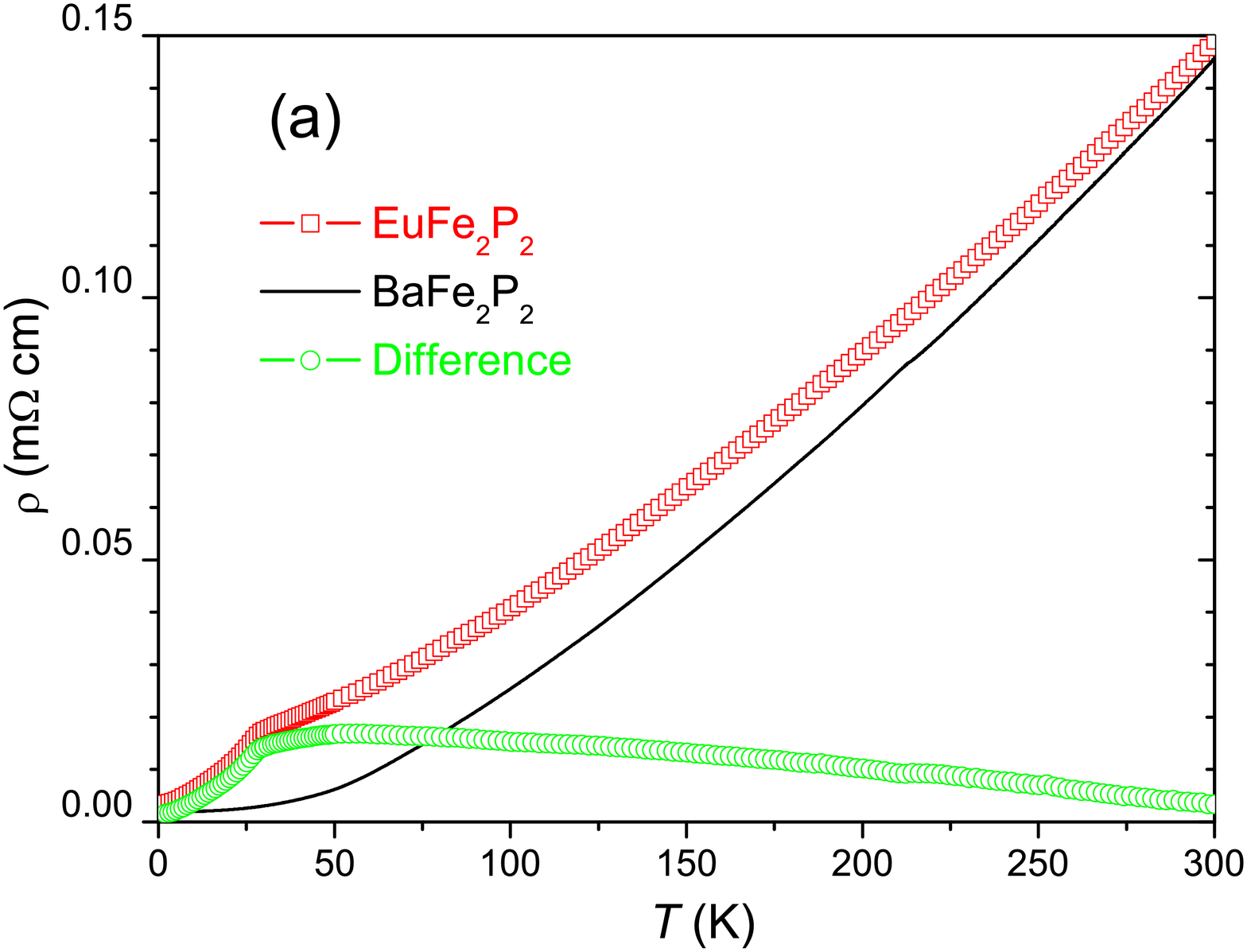}
\includegraphics[width=7cm]{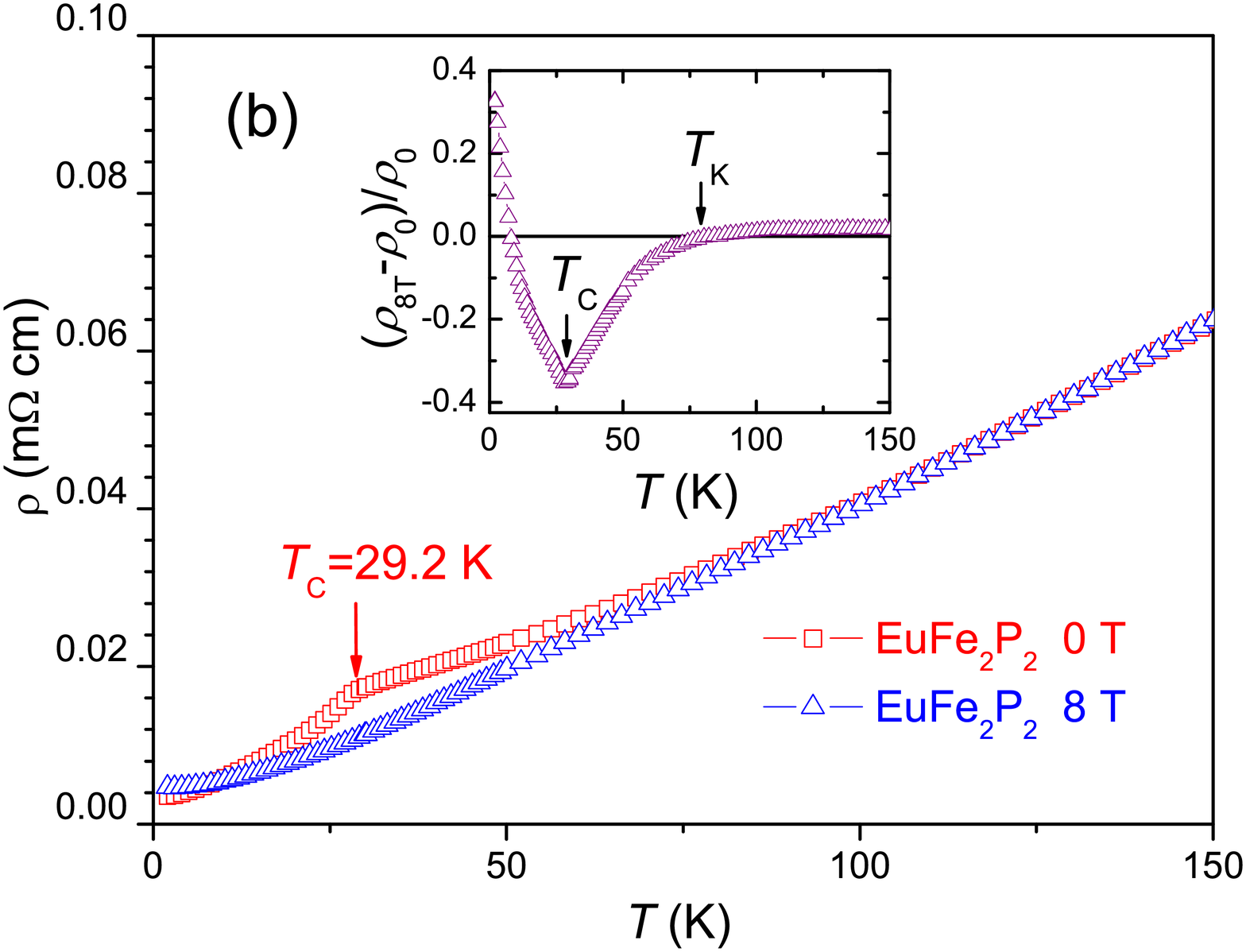}
\caption{(Color online) (a) Temperature dependence of resistivity
for EuFe$_2$P$_2$. The $\rho(T)$ data of BaFe$_2$P$_2$ are also
plotted for comparison. The green symbols are the difference (by a
subtraction), which basically represents the resistivity
contribution from magnetic scattering. (b) Anomalous temperature
dependence of magnetoresistance in EuFe$_2$P$_2$.}
\end{figure}

Under an 8 T field, there is negligible effect on $\rho(T)$ above 80
K. However, anomalous temperature-dependent MR is observed below 80
K, as illustrated clearly in figure 7(b). Namely, a negative MR
grows with decreasing temperature below 80 K and reaches its minimum
of $-$35\% at the FM ordering temperature. Then the negative MR
decreases with further decreasing temperature and finally undergoes
sign reversal around 10 K, below which positive MR increases with
decreasing temperature and achieves 35\% at 2 K. It is noted that
the resistivity kink at 29.2 K under zero field shifts to higher
temperature (over 50 K) and becomes very much broadened by the
external 8 T field. All the above MR behavior resembles those in
CeNiGe$_3$,\cite{PRB2003-Kondo} except that the latter system has an
AFM ground state.

The isothermal field dependence of MR for EuFe$_2$P$_2$ [figure
8(a)] gives further support for the dense Kondo behavior. For 40 K
$\leq$$T$$\leq$ 80 K, the negative MR increases monotonically with
increasing $H$ and decreasing $T$, in agreement with a
characteristic Kondo-type behavior. According to a theoretical
result,\cite{Schlottmann} The magnetoresistance, $\Delta \rho
/\rho_0$ can be scaled with $H/(T+T^*)$, where $T^*$ is a measure of
the single impurity Kondo energy scale. Figure 8(b) shows that the
MR data between 40 K and 80 K basically fall on the same curve for
$T^{*}=-$29 K. The negative sign of $T^{*}$, with the absolute value
close to the paramagnetic Curie temperature ($\theta$) derived
above, is consistent with the FM correlation in the system. This
scaling behavior provides compelling evidence for the Kondo
interaction in EuFe$_{2}$P$_{2}$.

\begin{figure}
\includegraphics[width=7cm]{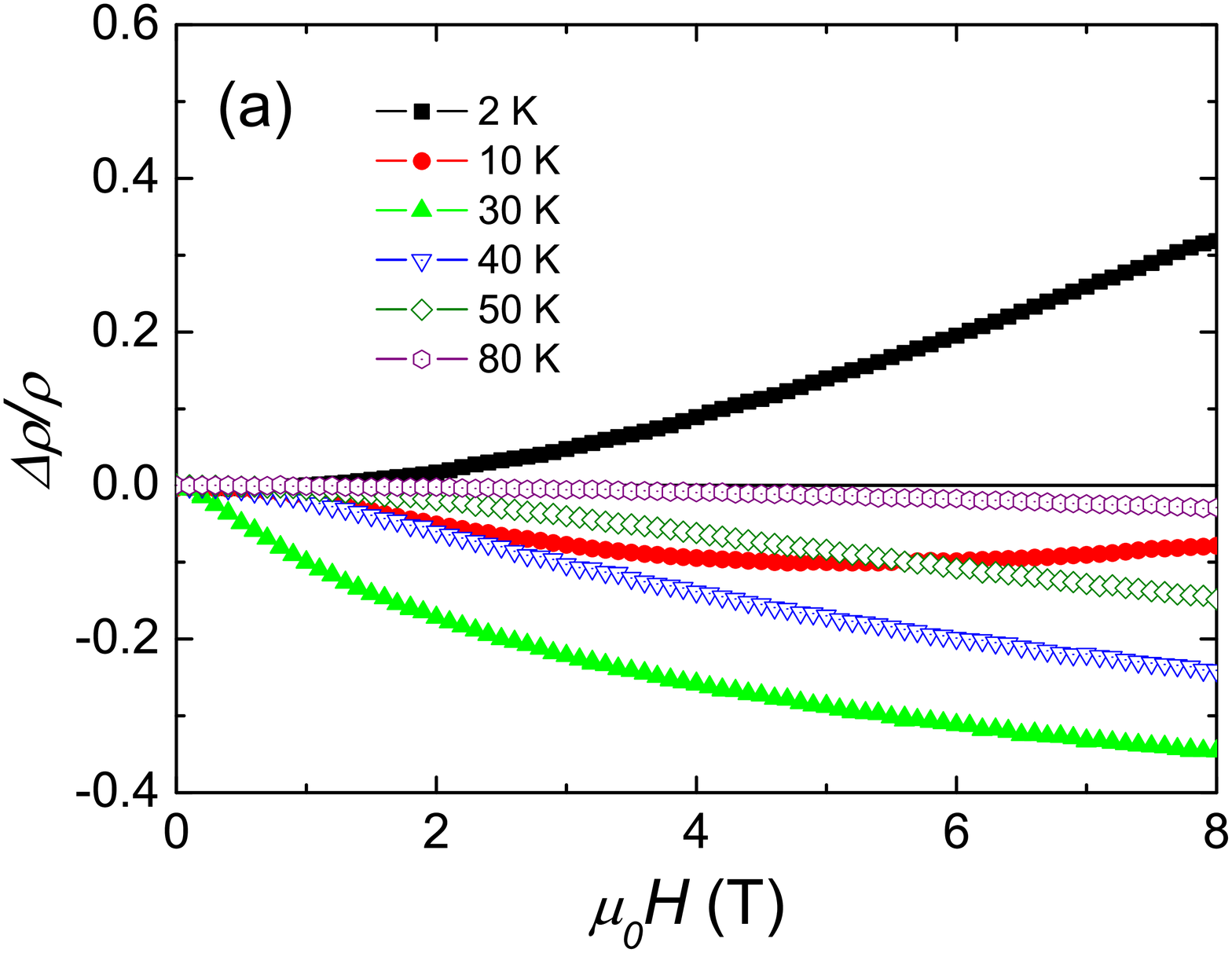}
\includegraphics[width=7cm]{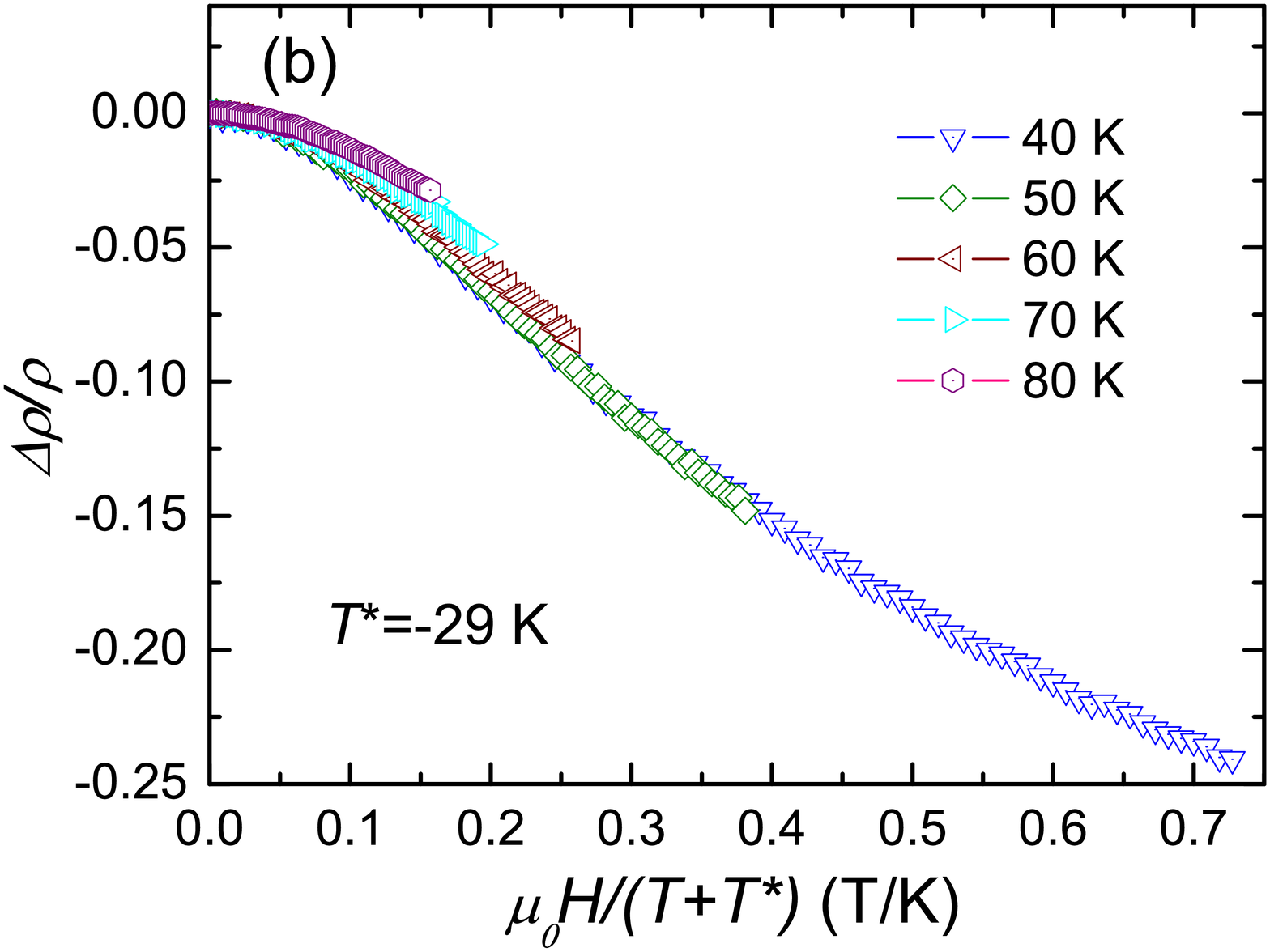}
\caption{(Color online) (a) Isothermal magnetoresistance for
EuFe$_2$P$_2$ at different temperatures. (b) Scaling behavior of the
magnetoresistance in EuFe$_2$P$_2$.}
\end{figure}

\subsubsection{\label{sec:level3}Thermoelectric Power}

Figure 9 shows the temperature dependence of thermoelectric power or
Seebeck coefficient ($S$) in EuFe$_2$P$_2$. The negative values of
$S$ in the whole temperature range indicate that electron transport
is dominant. This is in contrast with the $S(T)$ behavior in
EuFe$_2$As$_2$, which shows sign changes from negative to positive
then to negative again with increasing temperature.\cite{Ren2008}
Above 100 K, the thermopower is about $-10 \mu$ V/K, almost
independent of temperature (Note that the small gradual change from
200 to 300 K may due to the influence of trace amount of Fe$_2$P).
Below 90 K, $|S|$ starts to decrease, and it shows a linear behavior
from 15 to 45 K. According to Behnia {\it et al},\cite{Behnia2004}
the slope d$S$/d$T$ correlates closely with the electronic specific
heat coefficient $\gamma$ by a dimensionless quantity,

\begin{equation}
q=\frac{S}{T}\frac{N_{A}e}{\gamma},
\end{equation}

where $N_{A}e$ is the so-called Faraday number. For strongly
correlated electron systems, the $q$ value is close to
unity.\cite{Behnia2004} Thus the electronic specific heat
coefficient in EuFe$_2$P$_2$ can be estimated to be $\sim220$ mJ
K$^{-1}$ mol$^{-1}$, which is very close to the $C/T$ value at 2 K
(see below). This remarkably large value of $\gamma$ is consistent
with the Kondo behavior shown above.

\begin{figure}
\includegraphics[width=7cm]{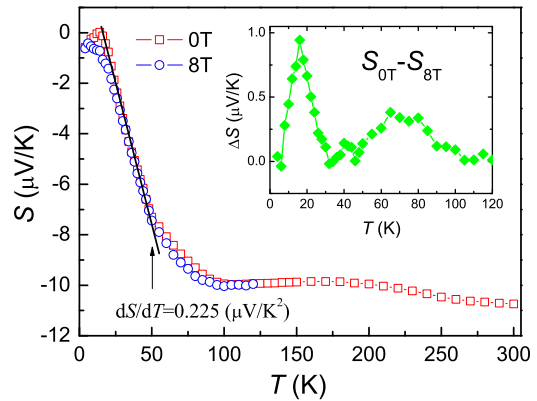}
\caption{(Color online) Temperature dependence of thermoelectric
power for EuFe$_2$P$_2$. The inset shows the change in thermopower
below the Curie temperature under an 8 T field.}
\end{figure}

Under a magnetic field of 8 T, the thermopower has a subtle change.
By a simple subtraction, one can see a two-peak structure in the
inset of figure 9. While the sharp peak at low temperatures is
related to the FM state, the broad peak centered at 70 K should be
associated with the dense Kondo effect, which is supposed to be
suppressed by the external field. Further theoretical investigation
is needed to clarify this phenomenon.

\subsection{\label{sec:level2}Specific Heat}

Figure 10 shows the specific heat measurement result for
EuFe$_2$P$_2$, especially in the lower temperature ranges. Under
zero field, a specific heat anomaly appears below 29 K,
corresponding to the FM and HM transitions of Eu$^{2+}$ sublattice.
The HM transition at $T_{HM}$=26 K is inferred by comparing the
$C(T)$ behavior with that of EuFe$_2$As$_2$, as shown in the upper
inset of figure 10. While EuFe$_2$As$_2$ shows a sharp peak at 19 K,
EuFe$_2$P$_2$ exhibits a round broader peak at 27 K, suggesting a
superposition of two nearby transitions. The successive transition
is further demonstrated by the decrease in $T_{HM}$ and an increase
in $T_C$ with the applied field of 0.2 T. The released magnetic
entropy up to the Curie temperature was estimated (the phonon
background is assumed to obey Debye model) about 80\% of $R$ln8
(where $R$ represents gas constant), implying the contribution from
Kondo state at high temperatures.

\begin{figure}
\includegraphics[width=7cm]{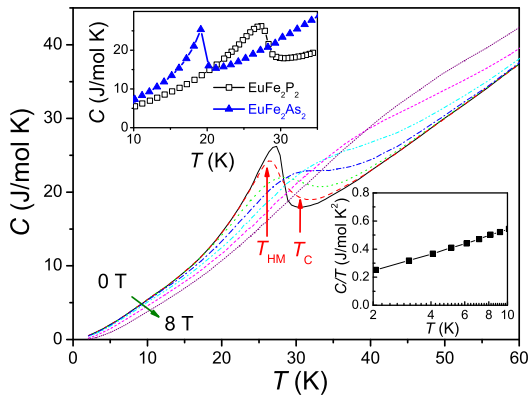}
\caption{(Color online) Temperature dependence of specific heat of
EuFe$_2$P$_2$ under magnetic fields of 0, 0.2, 0.5, 1, 2, 4 and 8
Tesla. The upper inset compares the $C(T)$ behavior with that of
EuFe$_2$As$_2$. The lower inset plots $C/T$ vs. $T$ for
EuFe$_2$P$_2$.}
\end{figure}

Owing to the FM/HM transitions, it is difficult to extract the
electronic specific heat coefficient by a conventional method. Thus
we simply consider $C/T$, shown in the lower inset of figure 10.
Since the phonon and magnon contributions are estimated to be
negligibly small at 2 K (since $\theta_D\gg$2 K and $T_C\gg$2 K),
the real electronic specific heat coefficient would be not so much
smaller than $C/T\mid_{T=2K}\approx$250 mJ K$^{-1}$ mol$^{-1}$. This
result is consistent the above indirect estimation from the
thermopower measurement.

Under magnetic fields, the specific heat anomaly moves to higher
temperatures and, the peak becomes more and more broadened with
increasing fields. This result is consistent with the above
magnetization measurement, suggesting dominant FM alignment and
stronger FM correlations in EuFe$_2$P$_2$, in comparison with the
$A$-type antiferromagnet EuFe$_2$As$_2$.\cite{Jiang2009,RXS,neutron}

\subsection{\label{sec:level2}Further Discussion}

Now let us discuss why the sister compounds EuFe$_{2}$P$_{2}$ and
EuFe$_{2}$As$_{2}$ behave so differently. First, according to the
argument by Si {\it et al},\cite{Si} the loss of Fe moments is due
to the relatively weak 3$d$ electron correlation in iron
\emph{phosphides}. So far, there is no report on the appearance of
Fe local moments in the iron phosphides with ThCr$_2$Si$_2$-type
structure. Second, the difference in the magnetic order of Eu 4$f$
moments has to be explained in terms of an indirect RKKY
interactions because the Eu-interlayer spacing is much larger than
the expected spacing for a direct exchange. The RKKY exchange
coupling $J_{RKKY}$ $\propto$ $-\frac{\alpha cos \alpha-sin
\alpha}{\alpha^{4}}$, where $\alpha$=2$k_{F}R$, \emph{R} denotes the
distance between two magnetic moments and $k_{F}$ the Fermi
wave-vector. Upon changing 2$k_{F}R$, $J_{RKKY}$ alters greatly, or
even changes the sign. In going from EuFe$_{2}$As$_{2}$ to
EuFe$_{2}$P$_{2}$, $R$ decreases from 6.06 to 5.64 {\AA}. Even if
$k_{F}$ remains constant, the decrease in $R$ may alter $J_{RKKY}$
remarkably, which would result in a crossover from AFM to FM
ordering. Note that the FM order of Eu$^{2+}$ moments in
EuFe$_{2-x}$Ni$_{x}$As$_{2}$ system was explained as a result of the
decrease in $\alpha$.\cite{Ren-PRB2009} In fact, the change in
$J_{RKKY}(R, k_{F})$ can be very minute, as manifested by the
helimagnetism in
Eu(Fe$_{0.89}$Co$_{0.11}$)$_2$As$_2$.\cite{Jiang-PRB2009} It is in
principle possible that successive magnetic transitions may occur
when decreasing temperature.

The observed dense Kondo behavior in EuFe$_{2}$P$_{2}$ could be
attributed to the proximity of the 4$f$ level to Fermi
energy.\cite{PRB2001} To our knowledge, dense Kondo behavior in
Eu-containing compounds is rarely discovered,\cite{EuCu2Si2}
primarily because Eu$^{2+}$ carries a large moment. Kondo effect
involves the intrasite coupling between local moment and conduction
carriers, while the RKKY interaction is a long-range intersite
magnetic exchange through conduction carriers. Therefore, the
interplay between Kondo and RKKY interactions is inevitable. In the
ground state of EuFe$_{2}$P$_{2}$, according to the well-known
Doniach scenario,\cite{Doniach} the RKKY interaction prevails
against the Kondo effect in EuFe$_{2}$P$_{2}$. Since applying
pressure enhances the Kondo effect,\cite{Uwatoko} the isovalent
chemical doping in EuFe$_{2}$P$_{2}$ should be promising to tune the
ground state in this intriguing system.

\section{\label{sec:level1}Concluding Remarks}

In summary, we have performed a systematic research on a ternary
iron phosphide EuFe$_2$P$_2$. This compound shows contrasting
physical properties with its analogue EuFe$_2$As$_2$, although both
materials contains Eu$^{2+}$ and without P-P covalent bonding. The
result indicates dominant FM ordering for the Eu sublattice.
However, the ground state has a possible helimagnetic ordering with
the moments basically parallel to the $c$ axis. Future neutron
diffractions are expected to resolve this issue. On the other hand,
the magnetotransport properties governed by the itinerant Fe 3$d$
electrons show a dense Kondo behavior. Future measurements using
single crystalline samples may confirm this point, and more
information on the anisotropic property are also expected.

Moreover, alloys of EuFe$_2$P$_2$ and EuFe$_2$As$_2$ exhibits
coexistence of high temperature superconductivity and local moment
ferromagnetism.\cite{Ren2009} Therefore, EuFe$_2$P$_2$ and its
related materials deserve further exploration with regard to the
interplay of Kondo, RKKY, and Cooper-pairing interactions among 4$f$
and conduction electrons.

\begin{acknowledgments}
This work is supported by the National Basic Research Program of
China (No. 2007CB925001), National Science Foundation of China (No.
10934005), and the Fundamental Research Funds for the Central
Universities of China. The research in Jerusalem is partially
supported by the Israel Science Foundation (ISF, Bikura 459/09), and
by the Klachky Foundation for Superconductivity.
\end{acknowledgments}


\begin{thebibliography}{00}

\bibitem{Radousky}H. B. Radousky,\emph{Magnetism in Heavy Fermion Systems}, (World
Scientific, Singapore, 2000).

\bibitem{Stewart}G. R. Stewart, Rev. Mod. Phys. \textbf{56}, 755 (1984).

\bibitem{JPCS1988}E. Morsen, B. D. Mosel, W. Muller-Warmuth, M. Reehuis, and W. Jeitschko, J. Phys. Chem. Solids
\textbf{49}, 85 (1988).
\bibitem{PB1998}C. Huhnt, W. Schlabitz, A. Wurth, A. Mewis, and M. Reehuis, Physica B \textbf{252},
44 (1998).

\bibitem{PRB2001}B. Ni, M. M. Abd-Elmeguid, H. Micklitz, J. P.
Sanchez, P. Vulliet, and D. Johrendt, Phys. Rev. B \textbf{63},
100102(R) (2001).

\bibitem{PRL1973}E. R. Bauminger, D. Froindlich, I. Nowik, S. Ofer, I. Felner, and I. Mayer, Phys. Rev. Lett. \textbf{30},
1053 (1973).
\bibitem{PRL1982}C. U. Segre, M. Croft, J. A. Hodges, V. Murgai, L. C. Gupta, and R. D. Parks, Phys. Rev. Lett. \textbf{49},
1947 (1982).
\bibitem{JSSC1978}R. Marchand, and W. Jeitschko, J. Solid State Chem. \textbf{24},
351 (1978).
\bibitem{Raffius}H. Raffius, E. M\"{o}rsen, B. D. Mosel, W. M\"{u}ller-Warmuth, W. Jeitschko, L. Terb\"{u}chte, and T. Vomhof, J. Phys. Chem. Solids \textbf{54},
135 (1993).
\bibitem{Ren2008}Z. Ren, Z. W. Zhu, S. Jiang, X. F. Xu, Q. Tao,
C. Wang, C. M. Feng, G. H. Cao, and Z. A. Xu, Phys. Rev. B
\textbf{78}, 052501 (2008).
\bibitem{JOP2008}M. Tegel, M. Rotter, V. Weib, F. M. Schappacher, R. Pottgen, and D. Johrendt, J. Phys: Condens. Matter \textbf{20},
452201 (2008).

\bibitem{Ren2009} Z. Ren, Q. Tao, S. Jiang, C. M. Feng, C. Wang, J. H. Dai, G. H. Cao, and
Z.-A. Xu, Phys. Rev. Lett. \textbf{102}, 137002 (2009).
\bibitem{Jeevan2008}H. S. Jeevan, Z. Hossain, D. Kasinathan, H. Rosner, C. Geibel,
and P. Gegenwart, Phys. Rev. B \textbf{78}, 092406 (2008).
\bibitem{Jiang2009}S. Jiang, Y. K. Luo, Z. Ren, Z. W. Zhu, C. Wang, X. F. Xu, Q. Tao, G. H. Cao, and Z.-A. Xu, New J. Phys. \textbf{11}, 025007
(2009).
\bibitem{Wu2009}D. Wu, N. Barisic, N. Drichko, S. Kaiser, A. Faridian, M. Dressel, S. Jiang, Z. Ren, L. J. Li, G. H. Cao, Z. A. Xu, H. S. Jeevan, and P. Gegenwart, Phys. Rev. B \textbf{79}, 155103 (2009).

\bibitem{hp1}C. F. Miclea, M. Nicklas, H. S. Jeevan, D. Kasinathan, Z. Hossain,
H. Rosner, P. Gegenwart, C. Geibel, and F. Steglich, Phys. Rev. B
\textbf{79}, 212509 (2009).
\bibitem{hp2}T. Terashima, M. Kimata, H. Satsukawa, A. Harada, K. Hazama, S. Uji, H. S. Suzuki, T. Matsumoto, and K. Murada, J. Phys. Soc. Jpn. \textbf{78}, 083701 (2009).

\bibitem{RXS}J. Herrero-Martin, V. Scagnoli, C. Mazzoli, Y. Su, R. Mittal, Y. Xiao, Th. Brueckel, N. Kumar, S. K. Dhar, A. Thamizhavel, and L.
Paolasini, Phys. Rev. B \textbf{80}, 134411 (2009).


\bibitem{neutron}Y. Xiao, Y. Su, M. Meven, R. Mittal, C. M. N. Kumar, T. Chatterji, S. Price, J. Persson, N. Kumar, S. K. Dhar, A. Thamizhavel, and Th. Brueckel, Phys. Rev. B \textbf{80}, 174424 (2009).

\bibitem{ChenGF} G. F. Chen, Z. Li, D. Wu, G. Li, W. Z. Hu, J. Dong, P. Zheng, J. L.
Luo, and N. L. Wang, Phys. Rev. Lett. \textbf{100}, 247002 (2008).
\bibitem{CeFePO}E. M. Bruning, C. Krellner, M. Baenitz, A. Jesche, F. Steglich, and C. Geibel, Phys. Rev. Lett. \textbf{101},
117206 (2008).
\bibitem{Rietan}F. Izumi, Mater. Sci. Forum \textbf{321-324}, 198
(2000).

\bibitem{ME-1}I. Nowik and I. Felner, Hyperfine Interactions \textbf{28}, 959 (1986).

\bibitem{JPC1985}R. Hoffman and C. Zhang, J. Phys. Chem. \textbf{89},
4175 (1985).
\bibitem{bvs1}I. D. Brown and D. Altermatt, Acta Cryst. \textbf{B41}, 244 (1985).
\bibitem{bvs2}N. E. Brese and M. O¡¯Keeffe, Acta Cryst. \textbf{47}, 192 (1991).

\bibitem{JPSJ1960}S. Chiba, J. Phys. Soc. Jpn. \textbf{15}, 581
(1960).

\bibitem{Jiang-PRB2009}S. Jiang, H. Xing, G. F. Xuan, Z. Ren, C. Wang, Z. A. Xu, and G. H. Cao, Phys. Rev. B \textbf{80},
184514 (2009).

\bibitem{Felner2009}I. Nowik and I. Felner, Physica C \textbf{469}, 485 (2009).
\bibitem{Ren-PRB2009}Z. Ren, X. Lin, Q. Tao, S. Jiang, Z. W. Zhu, C. Wang, G. H. Cao, and
Z.-A. Xu, Phys. Rev. B \textbf{79}, 094426 (2009).
\bibitem{112-Kondo}Y. Muro, N. Takeda, and M. Ishikawa, J. Alloys and Compounds
\textbf{257}, 23 (1997).

\bibitem{PRB2003-Kondo}A. P. Pikul, D. Kaczorowski, T. Plackowski, A. Czopnik, H. Michor, E. Bauer, G. Hilscher, P. Rogl, and
Yu Grin, Phys. Rev. B \textbf{67}, 224417 (2003).

\bibitem{Uwatoko}Y. Uwatoko and coworkers, Private Communications.
\bibitem{Thompson1985}J. D. Thompson and Z. Fisk, Phys. Rev. B
\textbf{31}, 389 (1985).
\bibitem{Schlottmann}P. Schlottmann, Phys. Rep. \textbf{181}, 1
(1989).
\bibitem{Behnia2004}K. Behnia, D. Jaccard, and J. Flouquet, J. Phys: Condens. Matter
\textbf{16}, 5187 (2004).

\bibitem{Si}Q. Si and E. Abrahams, Phys. Rev. Lett. \textbf{101},
076401 (2008).

\bibitem{EuCu2Si2}C. D. Cao, R. Klingeler, N. Leps, H. Vinzelberg, V. Kataev, F. Muranyi, N. Tristan, A. Teresiak, S. Q. Zhou, W. Loser, G. Behr, and B. Buchner, Phys. Rev. B \textbf{78},
064409 (2008).

\bibitem{Doniach}S. Doniach, \emph{Valence Instabilities and related narrow
band phenomena, edited by R. D. Parks,} (Plenum, New York, 1977).

\end{thebibliography}
\end{document}